\documentclass[11pt,preprint]{aastex}

\shorttitle{dSPH}
\shortauthors{Burkert}

\begin{document}

\title{The Structure and Dark Halo Core Properties of Dwarf Spheroidal Galaxies}

\author{A. Burkert\altaffilmark{1,2} }

\altaffiltext{1}{University Observatory Munich (USM), Scheinerstrasse 1, 81679 Munich, 
Germany}
\altaffiltext{2}{Max-Planck-Fellow,
Max-Planck-Institut f\"ur extraterrestrische Physik (MPE), Giessenbachstr. 1, 85748 Garching, Germany}

\email{burkert@usm.uni-muenchen.de}

\newcommand\msun{\rm M_{\odot}}
\newcommand\lsun{\rm L_{\odot}}
\newcommand\msunyr{\rm M_{\odot}\,yr^{-1}}
\newcommand\be{\begin{equation}}
\newcommand\en{\end{equation}}
\newcommand\cm{\rm cm}
\newcommand\kms{\rm{\, km \, s^{-1}}}
\newcommand\K{\rm K}
\newcommand\etal{{\rm et al}.\ }
\newcommand\sd{\partial}

\begin{abstract}
The structure and dark matter halo core properties of dwarf spheroidal galaxies (dSphs) are investigated.
A double-isothermal (DIS) model of an isothermal, non self-gravitating stellar system, in gravitational equilibrium and embedded 
in an isothermal dark halo core provides an excellent fit to 
the various observed stellar surface density distributions $\Sigma_*(r)$. Despite its constant velocity 
dispersion, the stellar system can be well characterised by King profiles \citep{king66} with a broad distribution
of concentration parameters $c=\log (r_{*,t}/r_{*,c})$, with $r_{*,t}$ and $r_{*,c}$ the stellar tidal and
core radius, respectively. The DIS model confirms the suggestion of
\citet{kormendy14} that the core scale length of the stellar system, defined as 
$a_* = - (d \ln \Sigma_*/d r^2)^{-1/2}$,
is sensitive to the central dark matter density $\rho_{0}$. In contrast to single-component systems,
$r_{*,t}$ however does not trace the tidal radius of the galaxy but the core radius $r_c$ of its dark matter halo.
$c$ is therefore sensitive to the ratio $\sigma_*/\sigma_0$ with $\sigma_*$ and $\sigma_0$ the stellar
and dark matter velocity dispersion, respectively. Simple empirical relationships are derived that allow
to calculate the dark halo core parameters $\rho_0$, $r_c$ and $\sigma_0$, given the observable quantities
$\sigma_*$, $a_*$ and $c$. The DIS model is applied to the Milky Way's dSphs. 
Their halo velocity dispersions lie in a narrow range of 10 km/s $\leq \sigma_0 \leq$ 18 km/s 
with halo core radii of 280 pc $\leq r_c \leq$ 1.3 kpc and $r_c \approx 2 a_*$.
All dSphs follow closely the same universal scaling relations $\langle \rho_0 r_c \rangle \equiv \rho_0 \times r_c = 75_{-45}^{+85}$ 
M$_{\odot}$ pc$^{-2}$ and $\sigma_0^2 \times r_c^{-1} = 0.45_{-0.27}^{+0.51}$ (km/s)$^2$ pc$^{-1}$
that characterise the cores of more massive galaxies over a range of 18 magnitudes in blue magnitude $M_B$. 
For given $\langle \rho_0r_c \rangle$ the core mass is a strong function of core radius, $M_c \sim r_c^2$. 
Inside a fixed radius $r_u$, with $r_u$ the logarithmic mean of the dSph's core radii,
the total mass $M_u = 2.17 \langle \rho_0 r_c \rangle r_u^2$ is however roughly constant.
Outliers with smaller masses are expected for dSphs with core radii that are much larger or smaller than $r_u$.
For the Milky Way's dSphs we find $r_u = 400 \pm 100$ pc and $M_u = 2.6 \pm 1.4 \times 10^7 M_{\odot}$,
in agreement with \citet{strigari08}.
Due to their small $r_c$, the core densities of the Galaxy's dSphs are very  higher,
with $\rho_0$ = 0.03 - 0.3 M$_{\odot}$ pc$^{-3}$. The dSphs would have to be
on galactic orbits with pericenters smaller than a few kpc in order for their stellar systems
to be affected by Galactic tides which is very unlikely. dSphs should therefore be tidally undisturbed.
Observational evidence for tidal effects might then provide a serious challenge for the cold dark matter scenario.
\end{abstract}


\section{Introduction}
The nature of dark matter is still a mystery. Standard cosmology works with the
assumption of a massive, weakly and gravitationally interacting particle \citep[e.g.][]{white82,steigman85,jungman96}. This 
cold dark matter (CDM) scenario
has proven to be very successful on large galactic and extragalactic scales, from cosmic
structure formation to the outer rotation curves of galaxies and the stability of
galactic disks \citep{ostriker73}. The success of the CDM model however is also to some extent frustrating
as any additional physical properties of the CDM particle remain hidden. Scientists 
are therefore searching for failures of CDM predictions that might lead to new insight into the
nature and origin of the dark matter particle.

Several small-scale problems of the CDM model have been discussed in the literature. These include the distribution of
satellite galaxies in large planar structures \citep[e.g.][]{kroupa05,ibata13, goerdt14} or the mass-luminosity problem of satellite galaxies
\citep{kroupa10}.
One of the most prominent heavily debated questions is the cusp-core problem \citep[e.g.][]{moore94,flores94,burkert95,strigari08,primack09,deblok10,
boylan11,ogiya14,ogiya14a,ogiya15}.  While simulations
predict cuspy central density profiles of CDM halos with the density increasing steeply towards the center
\citep{dubinski91,navarro97,moore99,dekel03a}, observations often indicate a flat 
dark matter density core \citep[e.g.][]{flores94,moore94,burkert95,deblok08,gentile09,oh11}. 
{ The detection of cored dark matter halos is not necessarily inconsistent with CDM. Various mechanisms
have been identified that can generate cores from an initially cuspy density distribution. Prominent examples are
fluctuations in the galactic potential, induced by
AGN feedback and galactic winds \citep[e.g.][]{navarro96,ogiya11,ogiya14,pontzen12,teyssier13,amorisco14} or gaseous and stellar clumps
spiraling to the center \citep[e.g.][]{elzant01,ma04,tonini06,goerdt10,inoue11}}

The cusp-core problem is best documented in low-mass dwarf galaxies
which are characterised by low baryon fractions and which therefore are ideal tracers of the underlying
dark halo structure, unperturbed by the gravitational influence of the baryonic component.
In addition, dwarf galaxies appear to host the highest density dark matter cores,
making an analysis of the halo structure easier. Ideally one would like to investigate 
H$\alpha$/HI rotation curves \citep[e.g.][]{carignan85,deblok01,corbelli14} which are a clear
tracer of the gravitational potential as function of radius. One of the problems however is
that for low-mass galaxies the stellar- and gas velocity dispersions begin to exceed their rotational velocity.
This is especially true for one of the smallest galaxies, known as dwarf spheroidals (dSph), 
the target of this paper. Recently, \citet[KF14;][]{kormendy14} investigated the dark halo scaling laws
in late-type galaxies, including dSphs \citep[see also][]{burkert97,salucci12}. Interestingly, the stellar velocity dispersion
$\sigma_*$ of the dSphs is of order 8-10 km/s, very similar to the universal turbulent velocity
of the diffuse gas component in most low-redshift star forming disk galaxies \citep{dib06}. 
However, in contrast to more massive galaxies, the 
gravitational field in dSphs is small, generating rotation curves of order 
$v_{rot} \approx 10$ km/s (see section 4).
Even if the star forming gas would have had enough angular momentum to settle initially into 
a thin disk configuration, stellar feedback processes could easily
destroy the disk, ejecting gas in a wind and leading to a dispersion-dominated spheroidal stellar system
\citep{navarro96,maller02,teyssier13}. 

KF14 consider a spherically symmetric stellar system with constant $\sigma_*$, embedded in a spherically symmetric
dark matter core with constant density $\rho_{0,d}$. Solving the hydrostatic equation (see equation 3) and
assuming that the stellar system is not self-gravitating they find that the stellar density distribution 
should be a Gaussian

\begin{equation}
\rho_*(r) = \rho_{0,*} \times \exp \left( - \frac{r^2}{a^2_*} \right)
\end{equation}

\noindent with $\rho_{0,*} = \rho_*(r=0)$ the central stellar density and 

\begin{equation}
a_* = \left( \frac{3 \sigma_*^2}{2 \pi G \rho_{0,d}} \right)^{1/2}
\end{equation}

\noindent the scale length of the stellar system. The projected stellar surface density distribution $\Sigma_*$ in this
case is also a Gaussian with the same scale length. Remarkably, KF14 show that many dSphs follow a Gaussian 
surface density distribution better than the typical exponential profile, seen for galactic disks. 

KF14 however also find that some dSphs cannot be fitted by a Gaussian. In addition, also dSphs with inner
Gaussian slopes show deviations further out. Another problem is the fact that isothermal, isotropic dark matter 
cores in equilibrium cannot have precisely constant densities, requiring a more detailed investigation.
Finally, equation 2 does not provide any information about the dark halo core radii, velocity dispersions
and masses. KF14 shift
the observed stellar scale length $a_*$ and velocity dispersion $\sigma_*$ along lines of constant
$\rho_{0,*}$ onto the core scaling relations of more massive galaxies, in order to infer the halo core 
properties. It is however not clear whether dSphs should follow the same
core scaling relations as more massive galaxies.

In this paper we therefore have a more detailed look at the coupled kinematics of stars and dark matter in dSphs, relaxing
the assumption of a constant density dark matter core.
Section 2 solves the hydrostatic equation of two isothermal particle systems, coupled by their
joint gravitational field. We derive formulae on how to determine the dark matter core density and how
to shift the observed stellar velocity dispersion $\sigma_*$ and central stellar scale length $a_*$ in order 
to infer the dark halo velocity dispersion $\sigma_d$ and halo core radius $r_{c,d}$. 
Section 3 then focusses on deviations from Gaussian profiles and the origin
of King profiles in isothermal dSphs. This section demonstrates that the stellar King concentration parameter $c_*$ is tightly
related to the dark halo velocity dispersion $\sigma_d$ and by this also makes it possible to determine the halo core radius
and core mass. These analytical results are then applied to the Milky Way's system
of dSphs in section 4 to investigate their halo core properties. Section 5 discusses the conjecture of 
\citet{strigari08} that dSphs have a universal mass, of order $10^7$ M$_{\odot}$, within a scale radius of 300 pc.
Section 6 summarises the results and concludes.

\section{The structure of double-isothermal (DIS) particle systems}

We assume that the cores of dark matter halos in dSphs are isothermal and isotropic with a constant
velocity dispersion $\sigma_{0,d}$. This assumption certainly has to break down
at some point as otherwise the dark halo mass would diverge as $M_d(r) \sim r$. 
For investigations of the outer halo regions, the Burkert profile \citep{burkert95}
might therefore provide a better approximation. It combines an isothermal-like inner core
with the characteristic $r^{-3}$ outer density decline, seen in most CDM simulations \citep{navarro97}. 

Observations also indicate
that the stellar body of dSphs is characterised by an almost constant velocity dispersion $\sigma_*$,
well beyond the half-light radius \citep[KF14]{walker09,evans09,salucci12}. \citet{majewski13} recently reported a 
drop in $\sigma_*$ in the heart of the Sagittarius dwarf spheroidal galaxy. { A similar feature is seen in
Sculptor \citep{breddels14}, indicating that some dSphs might be more complex 2-component galaxies
\citep{amorisco11}.  The dense cores are} limited to the very center that is small compared
the half-light radius. A central decline in $\sigma_*$ might in fact be a characteristic property
of dSphs in general. Many of them
show a strong increase in stellar surface density in the very center, requiring a change in their kinematics, most 
likely a declining velocity dispersion, in order to be in hydrostatic equilibrium.
The origin of cold nuclei in dSphs is certainly an interesting and yet unsolved problem. Here, however, we are interested in the global
structure of dSphs { which to good approximation is observed to be isothermal}
and will neglect their cold hearts. In addition, in order to keep the number of free parameters to a minimum,
we assume that both, the stellar components and the dark matter cores are isotropic with negligible anisotropy effects
\citep{ciotti99,evans09,salucci12}.

Adopting a central density $\rho_d (r=0) = \rho_{0,d}$, the radial density distribution $\rho_d(r)$ of an isothermal halo
in hydrostatic equilibrium is determined by 

\begin{equation}
\sigma_d^2 \frac{d \ln \rho_d}{dr} = -\frac{G M_d(r)}{r^2}
\end{equation}

\noindent Here $\sigma_d$ is the dark matter velocity dispersion and $M_d(r)$ is the cumulative dark matter
mass inside radius $r$. The assumption of
virial equilibrium might not always be valid \citep{kroupa97}, especially for tidal
dwarf galaxies \citep{wetzstein07,ploeckinger15} and strongly tidally
affected dSphs like the Sagittarius dwarf spheroidal \citep{ibata94,kroupa97,yang14}. Deviations from hydrostatic equilibrium
might therefore help to identify those (mostly outer) regions of dwarf satellites, that are tidally interacting with the host galaxy.
The thick grey line in the upper left and right panels of figure 1 shows the density distribution 
and logarithmic slope $d \ln \rho_d/d \ln r$ , respectively, of a non-singular isothermal sphere with finite $\rho_{0,d}$. 
Inside the core radius that for an isothermal sphere is defined as 

\begin{equation}
r_{c,d}^2 = \frac{9 \sigma_{d}^2}{4 \pi G \rho_{0,d}}
\end{equation}

\noindent the dark matter density distribution is roughly flat.
Note however that $\rho_d(r)$ is not exactly constant. This is due to the fact that,
in equilibrium, the pressure gradient $\sigma_d^2 d \ln( \rho_d)/dr$ has
to balance the gravitational force $GM_d(r)/r^2$. One might consider this a negligible effect for $r < r_{c,d}$. 
It is however precisely this small density gradient that determines the density distribution of the
embedded stellar component.

Let us therefore now include a non self-gravitating, isothermal stellar component, as observed for dSphs. The stars then
represent direct tracers of the underlying dark matter potential. It is unlikely that dSphs started that way.
Obviously, the gas clouds from which the stars formed, were self-gravitating \citep{burkert13,nipoti14}. 
A low star formation efficiency,
combined with strong galactic winds and ram pressure stripping might however have removed most of the baryons
before they could condense into stars. This conclusion is also supported by the low metallicites of dSphs
\citep{dekel86}.
If the galactic outflow was violent enough, in addition to leaving behind a non self-gravitating stellar system,
it could also have reshaped an initially cuspy dark halo, generating a core \citep[e.g.][]{navarro96,teyssier13}.

As both, the stars and the dark matter particles
move inside the same joint gravitational potential their density distributions are coupled:

\begin{equation}
\sigma_*^2 \frac{d \ln \rho_*}{dr} = -\frac{G M_d(r)}{r^2} = \sigma^2_{d} \frac{d \ln \rho_d}{dr},
\end{equation}

\noindent Here $\rho_*(r)$ and $\sigma_*$ correspond to the density
distribution and velocity dispersion of the stellar component.  Equation 5 leads to 

\begin{equation}
\rho_*(r) = A \times \rho_d^{\kappa}(r)
\end{equation}

\noindent with

\begin{equation}
\kappa = \frac{\sigma_d^2}{\sigma_*^2}
\end{equation}

\noindent $A$ is a constant of integration that determines the total stellar mass.
Equation 6 shows that it is indeed the dark matter density gradient that determines $\rho_*(r)$:

\begin{equation}
\frac{d \ln \rho_*}{d \ln r} = \kappa \frac{d \ln \rho_d}{d \ln r} .
\end{equation}

\noindent The three dashed lines in the upper panels of figure 1 show $\rho_*(r)$
for embedded stellar systems of our DIS model with velocity dispersions 
$\sigma_*/\sigma_{d}$ of 0.7, 0.5 and 0.25, corresponding
to values $\kappa$ of 2, 4 and 16, respectively. For $\kappa > 1.5$ the stellar density distribution
decreases faster than $r^{-3}$ in the outer region. In the academic limit that the DIS model holds for all r,
the stellar system would have a finite mass, despite the fact that it is isothermal at all radii.

\begin{figure*}
\centering
\includegraphics[width=1.0\textwidth]{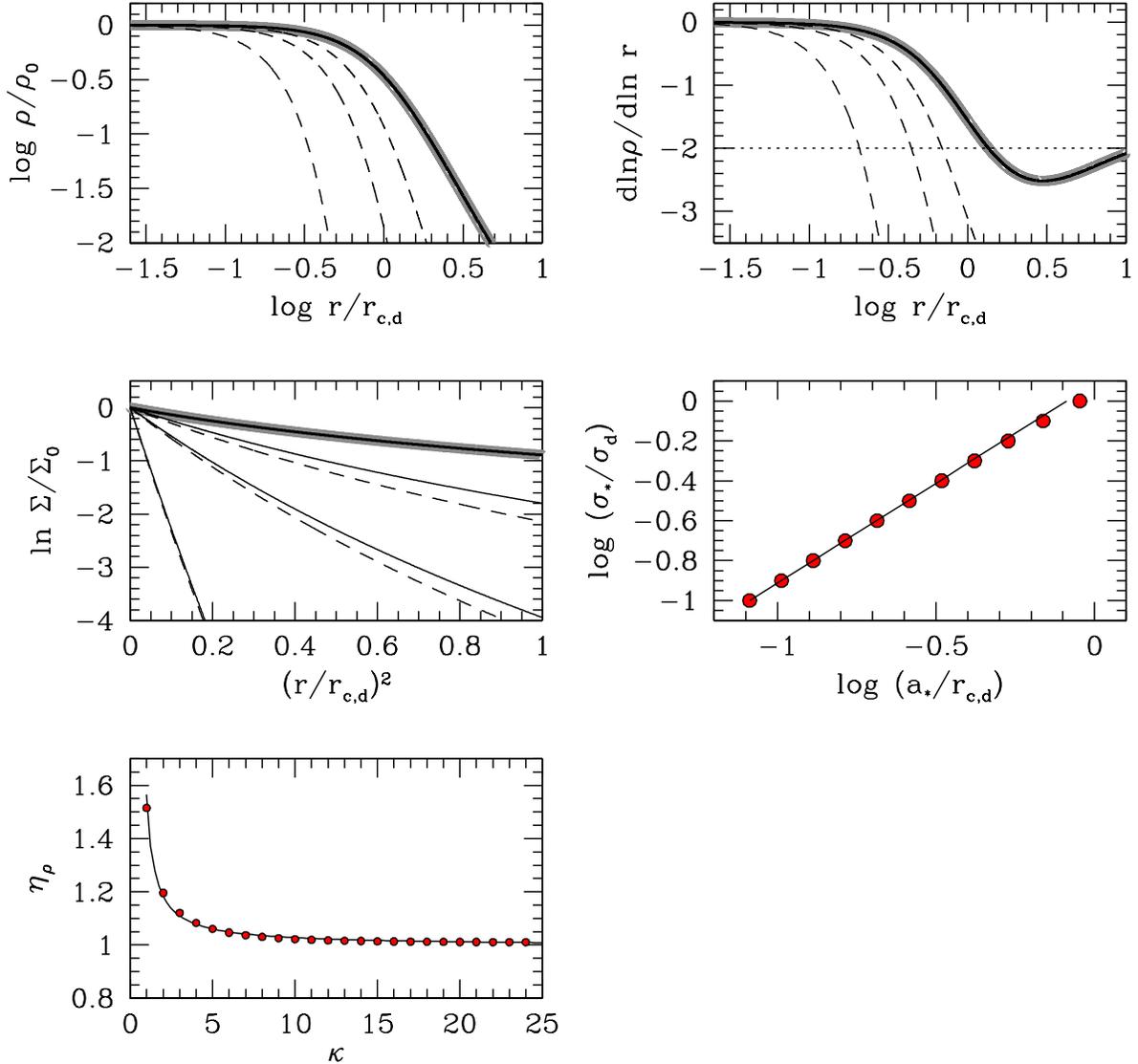}
\caption{
 \label{fig1}
The thick lines in the upper left and right panels show the normalized density distribution and logarithmic density slope
of an isothermal dark halo as function of radius.
The dashed lines depict three embedded stellar systems with $\sigma_*/\sigma_d = $ 0.7, 0.5 and 0.25, respectively. 
The thick line and the solid lines in the middle left panel show the surface density distribution
of the dark halo and its three embedded stellar systems, respectively. Dashed lines in the middle left panel 
correspond to the normalized
density distribution which is very similar to the surface density profile, especially for cold
stellar systems. The red points in the middle, right panel show the correlation between 
$\sigma_*/\sigma_d$ of dark matter confined stellar systems as function of $a_*/r_{c,d}$, as determined by integrating the
equations 3 and 6.  The solid line shows equation 9.
In the lower left panel the red dots show the correlation between the dark matter density parameter $\eta_{\rho}$
(equation 10) and $\kappa$, derived by an integration of the DIS model. The solid line represents the empirical
fit formula, equation 11.}
\end{figure*}

The middle, left panel of figure 1 shows the logarithm of the stellar density profile $\rho_*(r)$ and surface density
distribution $\Sigma_*(r)$ (dashed and solid lines, respectively) as function of $r^2$ for
$\kappa$ = 2,4, and 16. A Gaussian profile would
be represented by a straight line and indeed fits the inner profiles in general quite well.
The larger $\kappa$, the similar are the density and surface density distributions
and the more do the profiles resemble a Gaussian. 

Despite the fact that the dark matter core density is
not precisely constant and by that equation 2 is not exactly valid we can still formally
derive an approximate Gaussian scale length $a_* = - (d \ln \Sigma_*/dr^2)^{-1/2}$ 
by a least-squares linear fit of $\ln \Sigma_*$ versus $r^2$ within the innermost 
regions of the stellar component that we define as region, where $\Sigma_*$ decreases by a factor $e$ with respect
to the central value.  The middle right panel of figure 1 shows that $a_*/r_{c,d}$ depends
strongly on $\kappa$. Solving for $\rho_{0,d}$ in equation 2 and inserting it into equation 4 we find

\begin{equation}
\left( \frac{a_*}{r_{c,d}} \right) = 0.82 \left( \frac{\sigma_*}{\sigma_d} \right).
\end{equation}

\noindent It is not clear whether this should work, given the fact that equation 2 was derived
for a constant density core while we have argued that it is actually the dark matter density
gradient that determines the stellar density distribution (equation 8). However the
solid black line in figure 1 shows that equation 9 indeed provides an excellent fit to the actual data,
derived from a numerical integration of the DIS model (red points).

In the lower left panel of figure 1 we test the validity of equation (2) as an estimation
for the underlying dark halo density $\rho_{0,d}$, given $a_*$ and $\sigma_*$. If we write

\begin{equation}
\rho_{0,d} = \eta_{\rho} \frac{3 \sigma_*^2}{2 \pi G a_*^2}
\end{equation}

\noindent an analyses of the DIS model shows that $\eta_{\rho}$ depends only on $\kappa$ 
and in a way as shown by
the solid black line. For $\kappa \geq 1$ a very good approximation is (red points)

\begin{equation}
\eta_{\rho} \approx 1.01 (1 + 0.5 \exp [- (\kappa - 1)^{0.6}]).
\end{equation}

\noindent As expected, for very cold stellar systems with $\kappa \gg 1$ the stellar system
traces the innermost dark halo core with an almost constant density distribution. Here
$\eta_{\rho} \approx 1$ and equation 2 provides a good estimate of $\rho_{0,d}$. For kinematically
hotter stellar systems, however, equation 2 is not valid anymore and the correction factor
$\eta_{\rho}$ has to be taken into account.

\section{King profiles and apparent extra-tidal components}

Up to now we focussed on the innermost regions of dSphs that are sensitive 
to the central dark matter density. In order to gain information about the dark
halo core radii and core masses we need to look at stellar
traces further out. According to equation 6
very cold dSphs with $\kappa \gg 1$ populate regions that are deeply embedded in the dark halo core
and that are therefore not good probes to explore the larger environment.
One such example is Carina, shown in the upper left panel of figure 2. Carina can be
fitted well by a Gaussian over most of the stellar body with a small change in 
slope $d \ln \Sigma_*/dr^2$ in the outermost region.
In contrast, the stellar distribution of
the much more extended Sculptor dSph (upper right panel of figure 2) deviates strongly
from a Gaussian. The red solid lines in both figures show the surface density profiles
of DIS systems following equations 3 to 7 with 
$\kappa = 3.3$ and $\kappa = 1.5$ for Carina and Sculptor, respectively. With respect to the dark matter component,
the stellar system in Sculptor ($\sigma_* = 0.81 \sigma_d$) is kinematically hotter than in Carina 
($\sigma_* = 0.55 \sigma_d$), leading to a more extended structure.

\begin{figure*}
\centering
\includegraphics[width=1.0\textwidth]{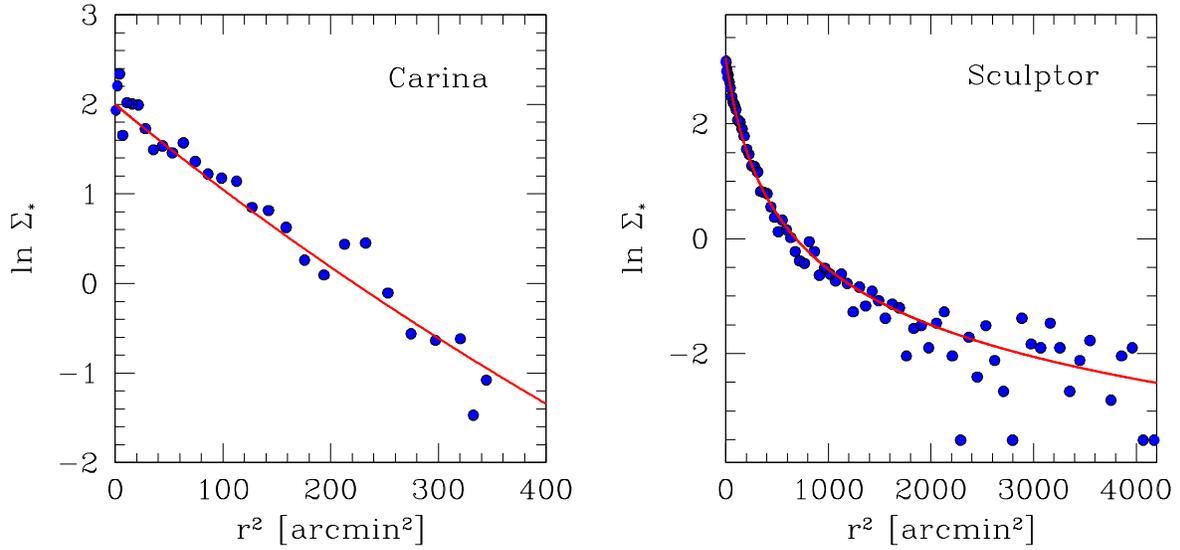}
\caption{
 \label{fig2}
The blue points in each panel show the stellar surface density measurements of \citet{irwin95} of 
the Milky Way dSphs Carina and Sculptor. { DIS models (red lines), normalized to the observed
central surface density and scale length, with $\kappa$ values as given 
by table 1 provide an excellent fit to the data.}}
\end{figure*}

It turns out that the equations 3 to 7 lead to surface density distributions that
fit all dSphs very well, even those with more complex, 
extended stellar components like Sculptor. This results from the fact
that dSphs are in general observed to follow King profiles \citep{king66,amorisco11} 
with different concentration parameters $c_*$.
{ King profiles are a 1-parameter family, characterised by the concentration parameter c which
is equal to the logarithm of the ratio of the outer cutoff radius to the core radius of a particle system}.
The same is true for the projected surface density distributions of stellar systems in our DIS model.
As an example, the solid lines in the left panel of figure 3 show
three different stellar systems of our DIS model with $\kappa$ values of 2.8, 1.4 and 1, respectively.
The points show the corresponding best fitting King profiles which have concentrations of 
0.6, 1.25 and 3.6, respectively.  Note that this excellent fit hides a fundamental difference 
between DIS models and King models. Our stellar systems have constant 
velocity dispersions. They are part of a 2-component system, with the stars being embedded in a surrounding dark halo
that has in general a different velocity dispersion than the stars. King models, instead, are one-component, self-gravitating
particle systems that are sometimes also called truncated isothermal spheres. They are characterised by
a special velocity distribution function that has been designed to fit stellar systems like globular clusters 
with sharp outer edges, generated as a result of
tidal stripping. In order for such a sharp outer edge to exist, 
the velocity dispersion of the stars in the King model has to decrease with radius with 
$\sigma_* \rightarrow 0$ at the outer edge. Otherwise stars would be able to move beyond it.
It is therefore surprising and at first not necessarily expected
that the one-component King profiles with a completely different kinematics provide such a good fit to the stellar structure 
of our two-component DIS systems, over more than 4 orders of magnitude in $\ln \Sigma_*$. 

A characteristic property
of King models is that the surface density structure changes strongly for concentrations
$1.2 \leq c_* \leq 2$ from a core with a steeply decreasing outer edge to a more extended
structure. The DIS models follow this trend nicely with kinematically hotter stellar systems,
characterised by smaller values of $\kappa$, corresponding to  King models with larger $c_*$.
In the transition regime, however, the best fitting King profiles are somewhat
steeper than the stellar systems for $\ln \Sigma_*/\Sigma_{0,*} \leq -4$ (see the
$c_*=1.25$ profile in the left panel of figure 3). Interpreting
such an extended population of stars as extra-tidal, in this case, would be misleading.
These stars are still deeply embedded and strongly bound to their dark halo. On the other hand,
an extra-tidal component detected at that level in $\ln \Sigma_*/\Sigma_{0,*}$ 
in systems with concentrations $c_* < 1$ or $c_* > 2$ cannot be explained within the framework of our
model and therefore might indeed represent a separate hot halo or even an extra-tidal component.

\begin{figure*}
\centering
\includegraphics[width=1.0\textwidth]{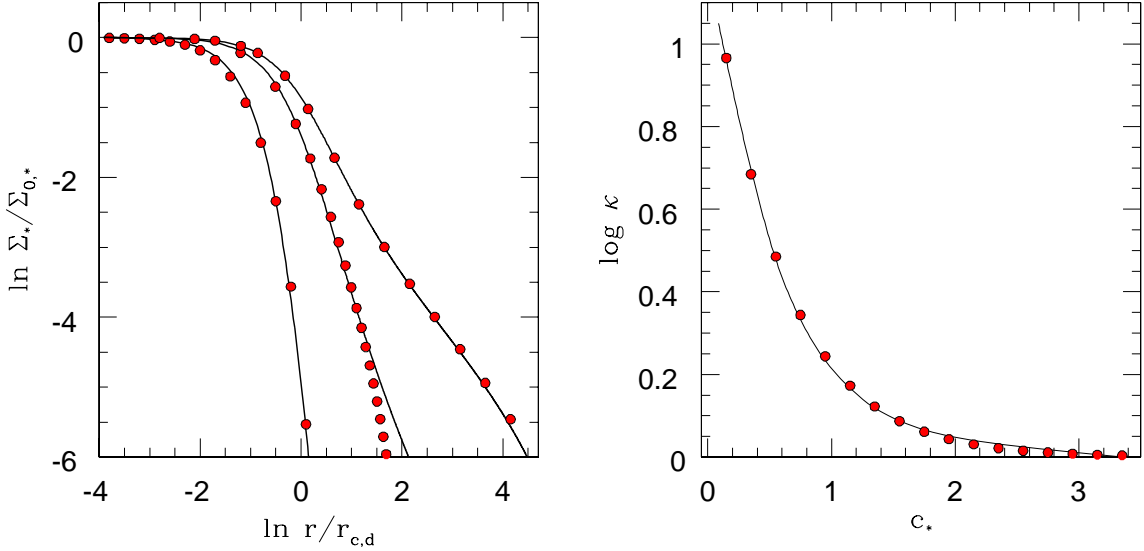}
\caption{
 \label{fig3}
The left panel compares the normalized surface density distribution of three characteristic King profiles 
(red points) with different concentrations $c_*=0.6$ (innermost contour), 1.25 (middle) and 3.6 (outer contour)
with the stellar surface density distribution 
of our DIS model (solid lines). Note the apparent "extra-tidal" stellar component for the
profile with $c_*=1.25$. The solid line in the right panel shows the correlation between the 
$\kappa$ values of the DIS model and the concentration of the best fitting King profile.
The red points show the empirical fit formula, equation 12. }
\end{figure*}

The right panel of figure 3 shows that there exists a tight correlation between the $\kappa$ value
of the DIS model and the King concentration parameter $c_*$ (solid black line).
The kinematically hotter the stellar system, i.e. the smaller $\kappa$, the 
more extended the stellar system and the larger $c_*$. The red points show the empirical relation

\begin{equation}
\log \kappa = 1.25 \exp (-1.72 c_*)
\end{equation}

\noindent which is an excellent fit to the data. 
Given $a_*$, $\sigma_*$ and $c_*$ one can now use the equations 9 to 12 and  calculate
the central density $\rho_{0,d}$, velocity dispersion $\sigma_{d}$ and core radius $r_{c,d}$ of the dark halo:

\begin{eqnarray}
\log \left( \frac{r_{c,d}}{\mathrm{pc}} \right) & = & 0.088+ \log \left( \frac{a_*}{\mathrm{pc}} \right) + 0.625 \exp (-1.72c_*) \nonumber \\
\log \left( \frac{\sigma_{d}}{\mathrm{km/s}} \right) & = & 0.625+ \log \left( \frac{\sigma_*}{\mathrm{km/s}} \right) + 0.625 \exp (-1.72c_*) \\
\log \left( \frac{\rho_{0,d}}{\mathrm{M}_{\odot}/\mathrm{pc}^3} \right) & = & 2.04+ 2 \log \left( \frac{\sigma_*}{\mathrm{km/s}} \right) -
 2 \log \left( \frac{a_*}{\mathrm{pc}} \right) \nonumber
\end{eqnarray}

\section{The dark halo core properties of local Milky Way dSphs}

\begin{table}
\begin{center}
\caption{Physical properties of the stellar and dark halo component
of the 8 classical Milky Way dSphs (KF14, IH95, McConnachie 2012).}

\vspace{0.3cm}

\begin{tabular}{cccc|ccccc}
\hline
 & \multicolumn{3}{c}{stellar component} & \multicolumn{5}{c}{dark matter component}\\
 & \multicolumn{1}{c}{a$_*$ [pc]} & \multicolumn{1}{c}{$\sigma_*$ [km/s]} &
   \multicolumn{1}{c|}{c$_*$} & \multicolumn{1}{c}{r$_{c,d}$ [pc]} &
   \multicolumn{1}{c}{$\sigma_d$ [km/s]} & \multicolumn{1}{c}{$\rho_{0,d}$ [M$_{\odot}$/pc$^3$]} & 
   \multicolumn{1}{c}{M$_{c,d}$ [10$^7$ M$_{\odot}$]} & \multicolumn{1}{c}{M$_{300}$ [10$^7$ M$_{\odot}$]}\\
\hline
Carina  & 202 & 6.6 & 0.51 & 450  & 12.0  & 0.13  & 2.6 & 0.9 \\
Draco   & 176 & 9.1 & 0.50 & 397  & 16.7  & 0.33  & 4.4 & 2.1 \\
Leo I   & 221 & 9.2 & 0.58 & 460  & 15.6  & 0.22  & 4.6 & 1.6 \\
Leo II  & 174 & 6.6 & 0.48 & 400  & 12.4  & 0.18  & 2.4 & 1.2 \\
UMi     & 211 & 9.5 & 0.51 & 470  & 17.3  & 0.25  & 5.6 & 1.9 \\
Fornax  & 705 &11.7 & 0.72 &1310  & 17.7  & 0.035 & 17.3& 0.4 \\
Sextans & 400 & 7.9 & 0.98 & 641  & 10.3  & 0.053 & 3.0 & 0.5 \\
Sculptor& 189 & 9.2 & 1.12 & 286  & 11.4  & 0.33  & 1.7 & 1.5 \\
\hline
\end{tabular}
\end{center}
\label{table:MWdsph}
\end{table}

As an application we investigate the dark halo core properties of the 8 classical
Milky Way dSphs (table 1), observed
by \citet[IH95, see also Kormendy \& Freeman 2014]{irwin95}. IH95 determined their stellar
surface density distribution with
high enough resolution in order to derive King concentration parameters and determine
the central scale lengths. IH95 also provide stellar velocity dispersions.
The central Gaussian scale lengths $a_{maj}$ were determined from 
major axis surface brightness profiles, shown in figure 2 of IH95.
These values are very close to the King core radii, summarized in Table 4 of IH95.
Following IH95, $a_*$ was then derived as the geometrical mean along
the major and minor axis with $a_* = a_{maj} \times r_{c,g,IH}/r_{c,IH}$ where
$r_{c,g,IH}$ and $r_{c,IH}$ is the stellar system's geometric mean and major axis core radius, respectively, as
determined by IH95.

\begin{figure*}
\centering
\includegraphics[width=1.0\textwidth]{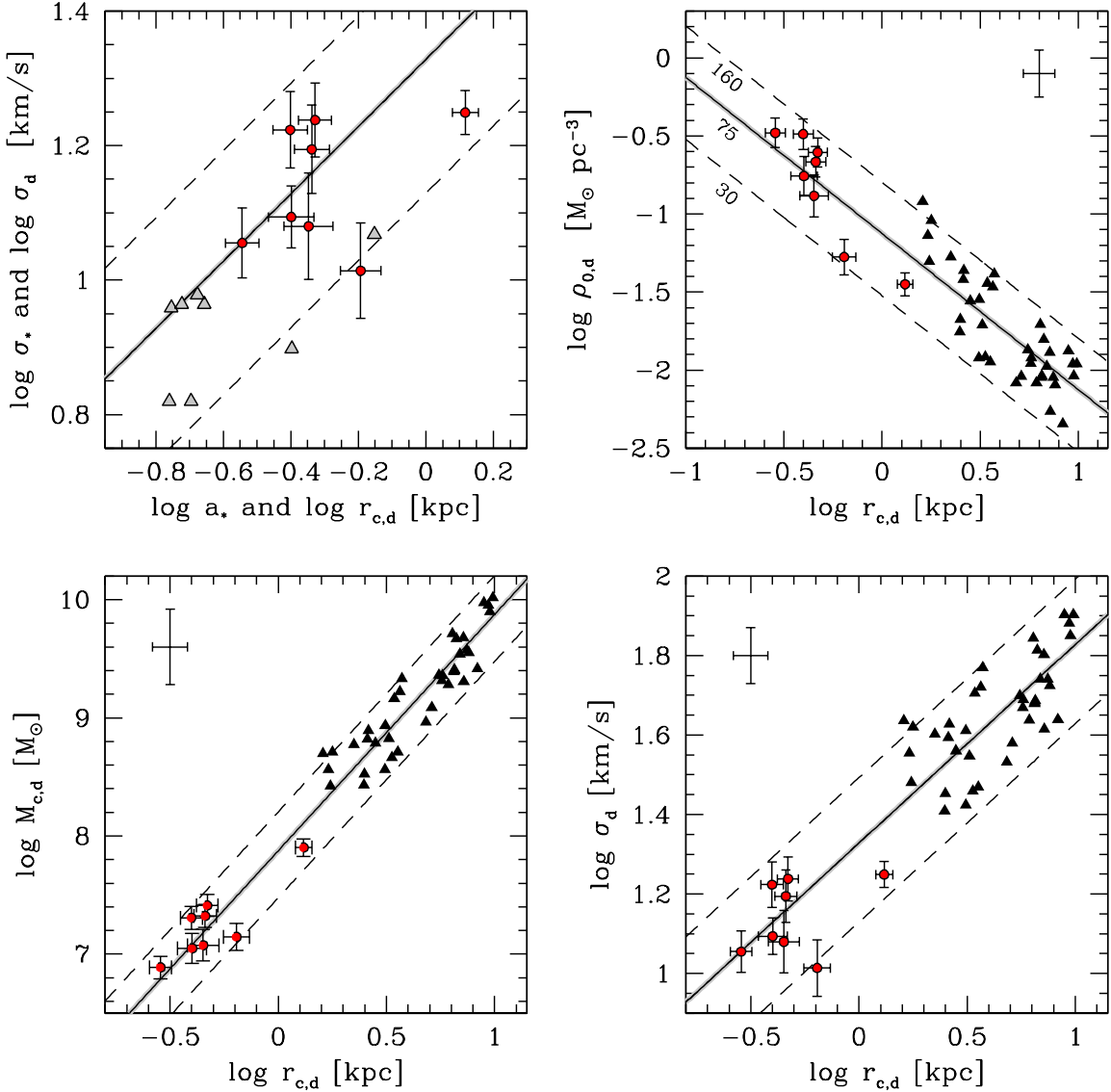}
\caption{
 \label{fig4}
The grey triangles in the upper left panel show the stellar velocity dispersion $\sigma_*$ versus the central stellar
scale length $a_*$ of the 8 classical Milky Way dSphs. 
Red points with errorbars depict the corresponding dark halo velocity dispersion $\sigma_{d}$
versus the halo core radius $r_{c,d}$. The red points in the upper right and lower left and right 
panels show the correlation  of $r_{c,d}$ with the dark halo central densities $\rho_{0,d}$, 
core masses $M_c$ and velocity dispersions $\sigma_d$, respectively.
Black triangles depict the core properties of more massive galaxies. { Typical error bars for these data points
are shown in the upper right or left corners of each plot.} The thick grey line in all
four panels corresponds to the dark halo core scaling relation $\langle \rho_{0,d} r_{c,d} \rangle$ = 75 M$_{\odot}$ pc$^{-2}$ 
that fits all galaxies very well.  The two dashed lines show the observed spread with the upper 
and lower limit corresponding to core surface densities of 30 and 160  M$_{\odot}$ pc$^{-2}$, respectively.}
\end{figure*}

\noindent Using the set of equations (13) we now can calculate the halo core parameters $r_{c,d}$, $\sigma_d$ and
$\rho_{0,d}$.  The results are summarized in table 1. 
The upper left panel of figure 4 shows $\sigma_*$ versus $a_*$
(grey triangles) and $\sigma_d$ versus $r_{c,d}$ (red points with error bars) for the 8 dSphs. 
{ Since IH95, updated stellar velocity dispersion measurements have been published
\citep[e.g.][]{walker09,mcconnachie12}. Here we adopt the values of $\sigma_*$ as given in the
regularly updated McConnochie data base \citep{mcconnachie12}}.
Dark halo cores are hotter than their stellar systems with velocity dispersions in between 10-18 km/s
and, on average, $\sigma_d \approx 1.6 \sigma_*$, corresponding to $\kappa \approx 2.56$. 
The halo core radii lie in the range of 290 pc to 1.3 kpc.
On average, $r_{c,d} \approx 1.9 a_*$. \citet{donato04} analysed a sample of high-resolution rotation curves of 25
disk galaxies and determined independently the disk scale lengths $r_{disk}$ and dark matter core radii. 
They found that both radii are strongly correlated with $r_{disk} \approx 2.4 r_{c,d}$. It is remarkable that
the ratio between the stellar scale length and the dark halo scale length is the same (of order 2) 
in very different galactic systems over a large range of mass. The origin is still unclear and might provide
further insight into the mechanisms that lead to dark matter cores.

The upper right panel of figure 4 shows the dark halo
central surface densities $\rho_{0,d}$ as function of $r_{c,d}$. Typical values are 0.2 M$_{\odot}$ pc$^{-3}$ with
a  range of 0.03 - 0.3 M$_{\odot}$ pc$^{-3}$.
It has been argued that dark halo cores follow a universal scaling relation with constant core surface density
$\langle \rho_{0,d} r_{c,d} \rangle$ \citep[e.g.][]{athanassoula87,burkert95,salucci00,kormendy04,deblok08,gentile09,donato09,
kormendy14,saburova14,cardone12}. KF14 find that this scaling relation holds over more than 18 magnitudes in M$_B$.
The black triangles in the upper right panel and in both lower panels of figure 4 show the core properties
of galaxies, compiled from
the literature by KF14 (see the list of references for the original data in table 1 of KF14). The core surface
densities of all galaxies lie in a narrow range of 
30 M$_{\odot}/$pc$^2 \leq  (\rho_{0,d} \times r_{c,d}) \leq$ 160 M$_{\odot}/$pc$^2$ (dashed lines). The 8 Milky Way
dSphs fall precisely into this regime, despite the fact that their $r_{c,d}$ are on average a factor 6 smaller
with $\rho_{0,d}$ being a factor of 6 larger. The lower left panel of figure 4 shows the dSph core masses, which for
non-singular isothermal spheres are 

\begin{equation}
M_{c,d} \equiv 2.17 \times \rho_{0,d} \ r_{c,d}^3 = 162.75 \left(\frac{(\rho_{0,d} \times r_{c,d})}{75 \mathrm{M}_{\odot} 
\mathrm{pc}^{-2}} \right) \left( \frac{r_{c,d}}{\mathrm{pc}} \right)^2 \mathrm{M}_{\odot}
\end{equation}

\noindent The core masses cover a range of one order of magnitude with masses in between (see table 1) 
$1.7 \times 10^7 \leq M_{c,d} \leq 1.7 \times 10^8$ M$_{\odot}$. 
The expected correlation between $M_{c,d}$ and $r_{c,d}$ (equation 14) is drawn for 
$(\rho_{0,d} \times r_{c,d}) = $ 30, 75 and 160 M$_{\odot}$ pc$^{-2}$,
together with the core masses of more massive galaxies. 
The dSphs follow the same core mass scaling relations as massive galaxies with the same spread.
Finally, the lower right panel shows again $\sigma_d$ versus $r_{c,d}$.
Now we compare the dSphs with the more massive galaxies. Both follow the same
universal scaling relation $\sigma_d^2 \times r_{c,d}^{-1} = 0.45_{-0.27}^{+0.51}$ (km/s)$^2$ pc$^{-1}$
\citep[]{devega14}, again with precisely the same spread.

\section{The origin of a common mass and length scale for dark matter cores}

\citet{strigari08} proposed that all dSphs of the Milky Way have the same total dark matter mass
$\log (M_{300,d}/M_{\odot}) = 7.0_{\ -0.4}^{\ +0.3}$, contained within a radius of 300 pc. 
The origin of a universal and constant mass might at first appear surprising, given the fact that the core masses
are a strong function of core size (equation 14). 
Note however that $M_{c,d}$ is measured within $r_{c,d}$ whereas M$_{300,d}$ is the mass
within a fixed radius 300 pc that can be smaller or larger than $r_{c,d}$.  
The question still arises why there should exist such a universal radius 
$r_{u,d}$, inside which halo cores have the same mass $M_{u,d}$ 
and what determines this radius. In addition,
adding Andromeda dSphs, \citet{collins14} find outliers that are not consistent
with the Strigari et al. mass which indicates that the situation can be more complex.

\citet{ogiya14a} discussed a possible connection between the existence of a universal mass scale
and the universal core surface density of dSphs. Following \citet{ogiya14a}, let us now explore this question within
the context of the DIS model. We start with a population of dSphs that has a common core surface
density $\langle \rho_{0,d} \times r_{c,d} \rangle$. Given $r_{c,d}$, one can determine $\rho_{0,d}$ and with equation (4) 
$\sigma_d$ . An integration of equation (3) then gives the complete density profile and by this $M_{u,d}$
for any given value of $r_{u,d}$.
The horizontal lines in figure 5 show $M_{300,d}$ as function of $r_{c,d}$ for the typical surface densities
$\langle \rho_{0,d} \times r_{c,d} \rangle = 75^{+75}_{-40}$ M$_{\odot}$ pc$^{-2}$ of the Milky Way dSphs.
Interestingly, $M_{300,d}$ is not continuously increasing with $r_{c,d}$ but instead has a maximum
at the point where it crosses the correlation between halo core mass and core radius (vertical lines).
This point also corresponds to the adopted characteristic scale length $r_{u,d} = 300$ pc.
The blue and red points in figure 5 show the dSph's $M_{300,d}$ and $M_{c,d}$, respectively. We added 10 dSphs from table 1 of 
\citet{strigari08} with given M$_{300,d}$ and King radii $r_{king,*}$ (blue triangles). The halo core radii are not known for this sample. 
For our joint sample we find on average $r_{c,d} \approx 2.3 \  r_{king,*}$ which was used in figure 5
in order to estimate $r_{c,d}$ for the dSphs with unknown core radii. The red points follow
the expected correlation between core mass and core radius (vertical solid and dotted lines).
The $M_{300,d}$ masses however show no such correlation with $r_{c,d}$ but instead are roughly constant.
Table 1 summarizes the $M_{300,d}$  values of our sample of Milky Way dSphs.
With on average $M_{300,d} = 1.3 \pm 0.6 \times 10^7$ M$_{\odot}$ they are in excellent 
agreement with \citet{strigari08}. 

The horizontal lines in figure 5 however show also that $M_{300,d}$ is not precisely constant but should
depend on the halo core radius.
This is true for any adopted radius $r_{u,d}$. Within the framework of the DIS model there exists no
universal radius $r_{u,d}$ inside which $M_{u,d}$ is constant for a population of dSphs with given core surface density.
It is however interesting that $M_{u,d}$  reaches a maximum for halos with core radii $r_{c,d}=r_{u,d}$. 
This is true for any adopted length scale $r_{u,d}$. As $M_{u,d}$ is very insensitive to $r_{c,d}$ in the flat vicinity
of this maximum, all dSphs with core radii in this regime would show very similar $M_{u,d}$ values, which is exactly
what we observe for the Milky Way's dSphs. The best choice of $r_{u,d}$ therefore is the average logarithmic core radius
of a given sample of dSphs with universal core surface densities. $M_{u,d}$ is then the mass within a dark halo
core with $r_{c,d} = r_{u,d}$ (equation 14):

\begin{eqnarray}
\log r_{u,d} = \frac{1}{N} \sum_{i=1}^N \log r_{c,d,i} \\
M_{u,d} = 2.17 \langle \rho_{0,d} \times r_{c,d} \rangle r_{u,d}^2 \nonumber
\end{eqnarray}

\noindent with N the number of dSphs and $r_{c,d,i}$ the core radius of galaxy i.
The origin of a maximum for $r_{c,d} = r_{u,d}$ can be easily understood. For $r_{c,d} > r_{u,d}$ the core radius is larger
than the region sampled by $r_{u,d}$ and the density is roughly constant with a value 
$\rho_{0,d} \sim 1/r_{c,d}$ due to the assumption of a constant core surface density. 
The enclosed mass with a given fixed radius r$_{u,d}$ is then $M_{u,d} \sim \rho_{0,d} \sim 1/r_{c,d}$ leading to $M_{u,d}$ decreasing with
increasing $r_{c,d}$.  For $r_{c,d} < r_{u,d}$ the 
region is larger than the core and extends out to radii where the dark matter density distribution begins to decrease.
To first order we can approximate $M_{u,d}$ now as the mass of a constant density core 
$\rho_d(r)=\rho_{0,d}$ for $r \leq r_{c,d}$ plus
the mass of an envelope with 
a power-law density distribution $\rho_d = \rho_{0,d} \times (r_{c,d}/r)^2$. As $\rho_{0,d} \sim 1/r_{c,d}$ we get
$M_{u,d} \sim r_{c,d}^2(r_{u,d}/r_{c,d} -2/3)$ which is a continuously increasing function with increasing $r_{c,d}$.

\begin{figure*}
\centering
\includegraphics[width=1.0\textwidth]{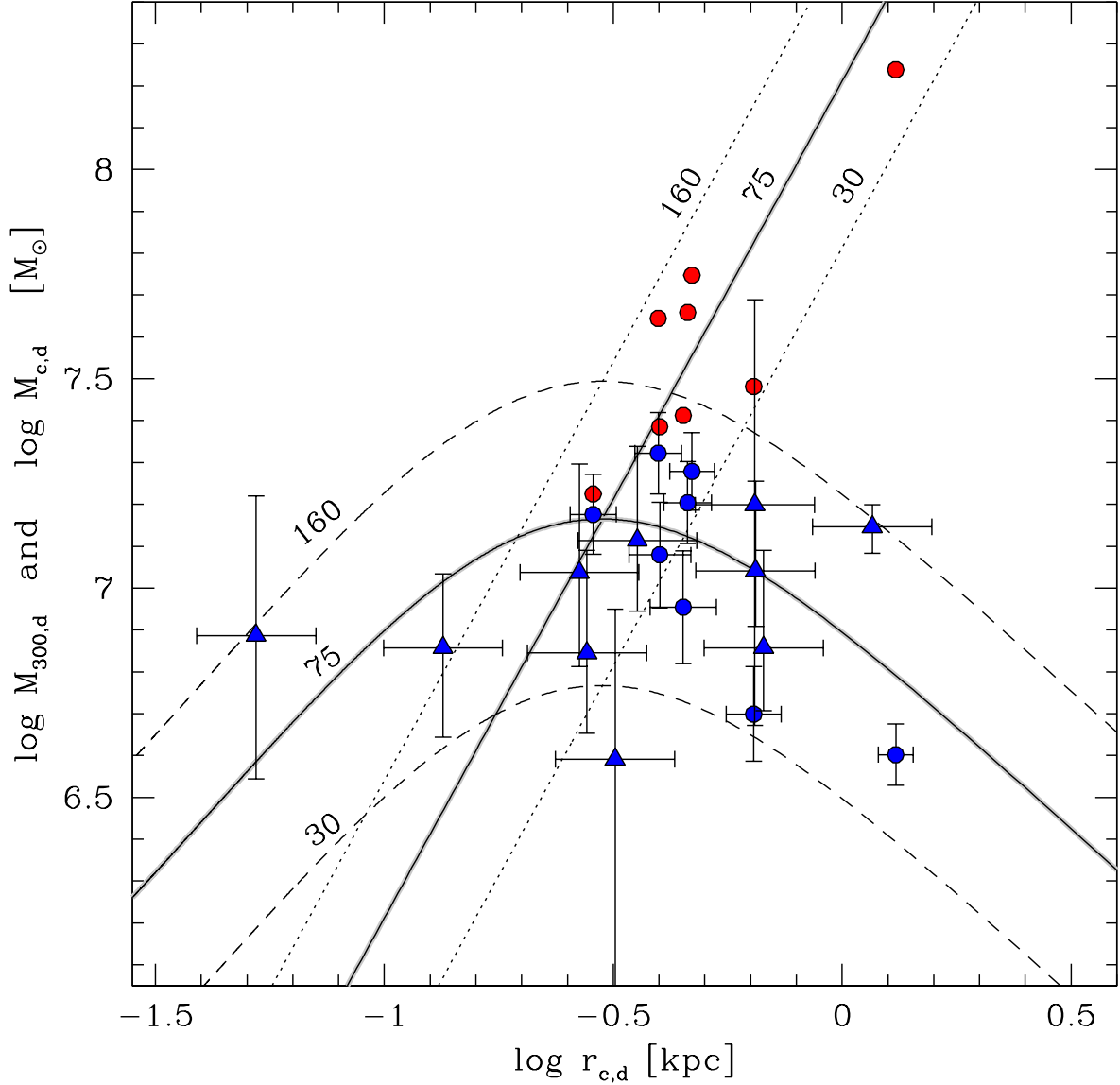}
\caption{
 \label{fig5}
Red and blue points show the dSph's core masses M$_{c,d}$ and masses 
M$_{300,d}$ within a fixed radius $r_{u,d}=300$ pc, summarized in table 1.
Blue triangles show additional dSphs taken from 
table 1 of \citet{strigari08}. The red points follow the core scaling relations 
$\rho_{0,d} \times r_{c,d} = 75^{+85}_{-40}$ M$_{\odot}$ pc$^{-2}$, observed for
massive galaxies (solid and dotted lines). The second set of solid and dashed lines
with a maximum at $r_{c,d} = 300$ pc shows the predicted relationship between $M_{300,d}$ and $r_{c,d}$
for halos with the same core surface densities. All lines are labeled according to their
core surface density in units of [M$_{\odot}$ pc$^{-2}$]. }
\end{figure*}

Applying equation 15 to our sample of dSphs we find 
$r_{u,d}  = 400 \pm 100$ pc and with $\langle \rho_{0,d} r_{c,d} \rangle$ = 75 M$_{\odot}$ pc$^{-2}$ get
$M_{u,d} = 2.4 \pm 1.4 \times 10^7$ M$_{\odot}$, in good agreement with \citet{strigari08}.
If the population has a large spread in $r_{c,d}$ we also expect to find 
outliers, populating the wings of the distribution further away from the maximum with smaller
masses $M_{u,d}$. This might explain the detection of outliers in Andromeda's system
of dSphs \citep{collins14}.  

\section{Summary and conclusions}

Motivated by the conjecture that dSphs are isothermal stellar systems \citep[e.g.][]{evans09}, 
embedded in isothermal
dark matter cores we investigated the structure of two particle systems with constant but different
velocity dispersions, in virial equilibrium within their joint gravitational potential. Note that here
the main objective was not to demonstrate that dark matter cores are isothermal and isotropic 
and it certainly has to break down outside of some radius \citep{burkert95}.
We worked with this assumption because it is the most simple model of a halo core
with the least number of free parameters. Of course, the fact that we find solutions 
that fit the observations very well is promising. But it is not a proof due to the fact that a fine tuned
radial distribution of anisotropy, coupled with a properly chosen gradient in velocity dispersion
could always lead to similar cored density profiles.

We demonstrate that the surface density distributions of the non self-gravitating stellar component of a
DIS galaxy can show a rich variety of profiles. They can formally be fitted by King profiles despite the fact 
that the DIS model consists of two isothermal components in contrast to the
one-component, non-isothermal King model. We find that the stellar systems in the DIS models 
have steeply decreasing outer edges, especially for high values of $\kappa$, 
not because they are tidally limited but because 
they are deeply embedded within the inner core regions of their dark halo.
This is in contrast to the real one-component King model which is characterised by a tidal radius,
the Roche radius, where the gravitational potential of the host galaxy begins to dominate and where
stars are unbound to the satellite. In addition,
DIS systems have projected stellar velocity dispersion profiles that remain constant all the 
way to their outermost radius. One-component King systems, on the other hand, show outer 
velocity dispersion profiles that decrease approaching $\sigma_* = 0$ at the tidal radius. 
Measurements of the stellar velocity dispersion of dSphs close to their cutoff radius 
therefore could help to distinguish tidally truncated one-component systems without confining dark halos
\citep{yang14} from those where the maximum radius is determined  by a strong dark matter confinement. 
This is also important, as the outer radii of satellite galaxies, interpreted as tidal radii, have been
used in order to gain information about the satellites' orbital parameters or the dark halo mass distribution 
of the host galaxy \citep[e.g.][]{pasetto11}. Our analyses instead shows that the maximum radii of the satellites
measure the core radii of their own dark halos and therefore do not contain information about the tidal radius
and the strength of the host galaxie's tidal field.

That $r_{t,*}$ traces the dark halo's core radii follows from the fact that the dSphs on average have
$\kappa \approx 2.6$. $r_{t,*}$ is close to the point, where the logarithmic density gradient
$d \ln \rho_*/d \ln r$ begins to decrease faster then -3 which, according to equation 8,
then corresponds to a dark halo density gradient of $d \ln \rho/d \ln r = -3/\kappa = -1.2$.
The upper right panel of figure 1 shows that this slope is close to the core radius $r_c$ of the dark
matter halo. It is not clear yet, whether this is a coincidence or whether the processes that generated
the dark halo cores and their non-self gravitating stellar tracer component naturally lead to such a
configuration \citep{dekel03b}.

For DIS systems, the central Gaussian scale length $a_*$, the velocity dispersion $\sigma_*$
and the concentration $c$ of the stellar component completely specify the dark halo core parameters
$\rho_{0,d}$, $\sigma_d$ and $r_{c,d}$. We determined
these parameters for 8 dSphs of the Milky Way and find that their dark halos have the same core surface densities
$\rho_{0,d} \times r_{c,d} = 75_{-45}^{+85}$ M$_{\odot}$ pc$^{-2}$ as more massive galaxies,  with exactly the same spread.
This is very puzzling as dSphs have a different structure and live in very different environments.
{ At the moment it is not clear whether this result is true also for M31's dSphs and whether it can be extended
to the ultra-faint satellite population  which due to their smaller radii should have even higher dark halo core densities.
An analyses similar to what was presented in this paper would require deeper observations of their stellar 
surface density distributions that are accurate enough in order to make King profile fits.}

The origin of dark matter cores is still not well understood. 
Suggestions range from gravitational interaction with the baryonic component 
\citep[e.g.][]{navarro96,elzant01,goerdt10,inoue11,
ogiya11,pontzen12,governato12,teyssier13,gritschneder13,ogiya14,ogiya14a} to a non-standard primordial power spectrum
\citep{zentner02,polisensky14}, warm dark matter \citep[e.g.][]{lovell14}, other intrinsic
properties of dark matter like self-interaction and self-annihilation \citep[e.g.][]{spergel00,burkert00,loeb11,elbert14}
or modifications of Newtonian dynamics \citep[e.g.][]{milgrom83,kroupa12}.
Whatever the mechanisms, considerable fine tune is required in order to generate a universal core
scaling relation over more than 18 orders of magnitudes in blue magnitude M$_B$ with exactly the same spread.

Adopting a constant core surface density, $M_{c,d}$ depends strongly on $r_c$. Focussing however on a fixed radius $r_{u,d}$, the
enclosed mass $M_{u,d}$ shows a different dependence on halo core radius, with a maximum at $r_{c,d} = r_{u,d}$. 
All halos with core radii in the vicinity of this maximum should therefore show similar values of $M_{u,d}$
which could explain the observations of \citet{strigari08}. The best choice of $r_{u,d}$ is therefore dependent on
the dSph's distribution of $r_{c,d}$. 
There does however not exist a universal mass scale $M_{u,d}$ that is independent of $r_{c,d}$. Smaller values of $M_{u,d}$
are expected for outliers with core radii that are very different from $r_{u,d}$. Turning this argument around,
if such a universal mass would exist, independent of $r_{c,d}$, it would be a clear signature that dark halo cores
are not isothermal.

As the core radii of dSphs are small, their core densities have to be high in order for the core surface density
to remain constant. This should shield dSphs efficiently against the tidal forces of their host galaxies.
Adopting a constant rotation curve $v_{rot}$, the Milky Way's mean density within a given radius r is

\begin{equation}
\langle \rho_{MW} \rangle = \frac{3 v_{rot}^2}{4 \pi G r^2} = 2.7 \left( \frac{v_{rot}}{220 \mathrm{km/s}} \right)^2
\left( \frac{\mathrm{kpc}}{r} \right)^2 \mathrm{M}_{\odot} \ \mathrm{pc}^{-3}
\end{equation}

\noindent The stellar system in dSphs would be tidally affected if 
$\langle \rho_{MW} \rangle > \rho_{0,d} \approx$ 0.2 M$_{\odot}$ pc$^{-3}$ which requires orbital 
pericenters of order a few kpc, which is very unlikely. dSphs therefore should be
strongly shielded from any tidal affects by their deep dark matter
potential wells and should survive as satellites of the Milky Way for a long time to come. 
However extra-tidal debris has been reported in some dSphs \citep[IH95][]{walcher03,majewski05,battaglia12}.
We discussed in section 3 that the DIS model in a certain concentration regime indeed leads 
to profiles that are somewhat more extended than the best fitting King profiles. 
This could be mis-interpreted as a tidal component. Strong evidence
for tidal interactions would however represent a real challenge for the existence of
a shielding dark halo, opening the door for alternative ideas
\citep{milgrom83,yang14}.

\acknowledgments

This work was supported by the cluster of excellence "Origin and Structure of the Universe". Thanks to the
referee for valuable comments that substantially improved the paper. I 
thank my CAST group at the University Observatory, Munich for inspiring discussions and suggestions.
Special thanks also go to Edvige Corbelli, Avishai Dekel, Ken Freeman, Pavel Kroupa, 
Go Ogiya, Jerry Ostriker, Paolo Salucci and David Spergel for interesting discussions and a 
careful reading of the manuscript. 
Finally I would like to thank the Harvard Center for Astrophysics
for support, lively discussions and an infinite supply of coffee that was essential and the driver for
finishing this paper.

\end{document}